\documentclass[preprint,longbibliography,superscriptaddress]{revtex4-1}
\usepackage{color}
\usepackage{graphicx}
\usepackage{mathrsfs}
\usepackage{bm}
\usepackage{hyperref}

\begin{document}

\title{Two-dimensional band structure in honeycomb metal-organic frameworks}

\author{Avijit Kumar}
\affiliation{Department of Applied Physics, Aalto University School of Science, PO Box 15100, 00076 Aalto, Finland}
\author{Kaustuv Banerjee}
\affiliation{Department of Applied Physics, Aalto University School of Science, PO Box 15100, 00076 Aalto, Finland}
\author{Adam S. Foster}
\affiliation{COMP Center of Excellence, Department of Applied Physics, Aalto University School of Science, P.O. Box 11100, FI-00076 Aalto, Finland}
\affiliation{Division of Electrical Engineering and Computer Science, Kanazawa University, Kanazawa 920-1192, Japan}
\affiliation{Institut f\"ur Physikalische Chemie,
Johannes Gutenberg-Universit\"at Mainz, Duesbergweg 10--14, D-55099 Mainz, Germany}
\author{Peter Liljeroth}
\email{peter.liljeroth@aalto.fi}
\affiliation{Department of Applied Physics, Aalto University School of Science, PO Box 15100, 00076 Aalto, Finland}

\maketitle

%\keywords{scanning tunneling microscopy (STM), metal-organic framework (MOF), cobalt, 4,4′-dicyanobiphenyl (DCBP), 9,10-Dicyanoanthracene (DCA), epitaxial graphene, Ir(111)}

\textbf{Metal-organic frameworks (MOFs) are an important class of materials that present intriguing opportunities in the fields of sensing, gas storage, catalysis, and optoelectronics \cite{Hendon:2017ACSCentSci, MOFsensors,MOFcatal,MOFopto}. Very recently, two-dimensional (2D) MOFs have been proposed as a flexible material platform for realizing exotic quantum phases including topological and anomalous quantum Hall insulators \cite{Wang2013,Wang2014,Zhang2016,Dong2016,Wang:2017APL,Zhang2017sc,Yamada2017ksl}. Experimentally, direct synthesis of 2D MOFs has been essentially confined to metal substrates \cite{Schlickum2007,Pawin2008_DCACu,Kambe2013,Urgel2015coord,Urgel2016,Stepanow2007,Dong:2016MOFreview,Sakamoto2017}
, where the interaction with the substrate masks the intrinsic electronic properties of the MOF. Here, we demonstrate synthesis of 2D honeycomb metal-organic frameworks on a weakly interacting epitaxial graphene substrate. Using low-temperature scanning tunneling microscopy (STM) and atomic force microscopy (AFM) complemented by density-functional theory (DFT) calculations, we show the formation of 2D band structure in the MOF decoupled from the substrate. These results open the experimental path towards %using MOFs as a testbed for 
MOF-based designer quantum materials with complex, engineered electronic structures with potential applications in devices with currently inaccessible and unforeseen functionalities \cite{Basov:2017quantummaterials}.}

The synthetic flexibility and tuneable electronic properties of MOFs stem from the choice of metal atoms, organic molecules, the linker chemistry and electronic and magnetic interactions among the building blocks \cite{ Barth:2010review,Dong:2016MOFreview,Wang2013,
Dong2016,Zhang2017sc,Yamada2017ksl}. For example, it is possible to realize honeycomb and Kagome lattices that are expected to give rise to Dirac cones and flat bands in the band structure \cite{Wang2013,Wang2014,Zhang2016,Dong2016}. Introducing spin-orbit coupling in hexagonal MOFs should result in the opening of topologically non-trivial band gaps and the realization of organic topological insulators. Further, the immense design flexibility suggests MOFs as an ideal tuneable platform for realizing organic quantum materials with exotic electronic ground states such as quantum anomalous Hall insulators, Kitaev spin liquids, and superconductors \cite{Wang2013,Wang2014,Zhang2016,Dong2016,Wang:2017APL,Zhang2017sc,Yamada2017ksl}. However, experimental study of these phases requires synthesis methods that yield monolayer MOFs on weakly interacting substrates such that their intrinsic electronic properties can be probed. 

Procedures for direct growth of 2D MOFs exist, \emph{e.g.} through synthesis on the air-liquid interface or by chemical vapour deposition (CVD) in ultra-high vacuum (UHV) conditions \cite{Kambe2013,Schlickum2007,Pawin2008_DCACu,Stepanow2007,
Dong:2016MOFreview,Sakamoto2017}.
CVD growth is typically carried out on metallic substrates where various types of frameworks have been studied in detail \cite{Barth:2010review,Dong:2016MOFreview}. However, the strong hybridization with the underlying substrate masks the intrinsic properties of the frameworks. This problem has been overcome in the case of single molecules by the use of ultrathin insulating films \cite{Qiu2003,Repp:2005PRL,Swart2011} and inert 2D materials such as graphene \cite{Jarvinen2013,liljeroth_jpcc_2014,Riss2014,Kumar2017review} that electronically decouple the molecule from the metallic substrate. Unfortunately, self-assembly and, in particular, on-surface chemical reactions are a virtual \emph{terra incognita} on weakly interacting, non-catalytic substrates \cite{Abel2011,Dienel2014,Morchutt2015,Guo2017,Schuller2016,Kumar2017review} and therefore the experimental observation of the intrinsic electronic properties of 2D MOFs has been elusive. 

Here, we demonstrate the controlled synthesis of high quality honeycomb MOFs on epitaxial graphene using different organic linkers (dicyanobiphenyl, DCBP, and dicyanoanthracene, DCA) with cobalt metal atoms. We characterize the structures using low-temperature STM and AFM. We demonstrate the formation of 2D band structure in the DCA-Co MOF by scanning tunneling spectroscopy (STS) measurements complemented by DFT calculations.

%\section{Results and discussion}

%\textbf{Structural Properties --} 

\begin{figure}[h]
  \includegraphics[width=0.8\textwidth]{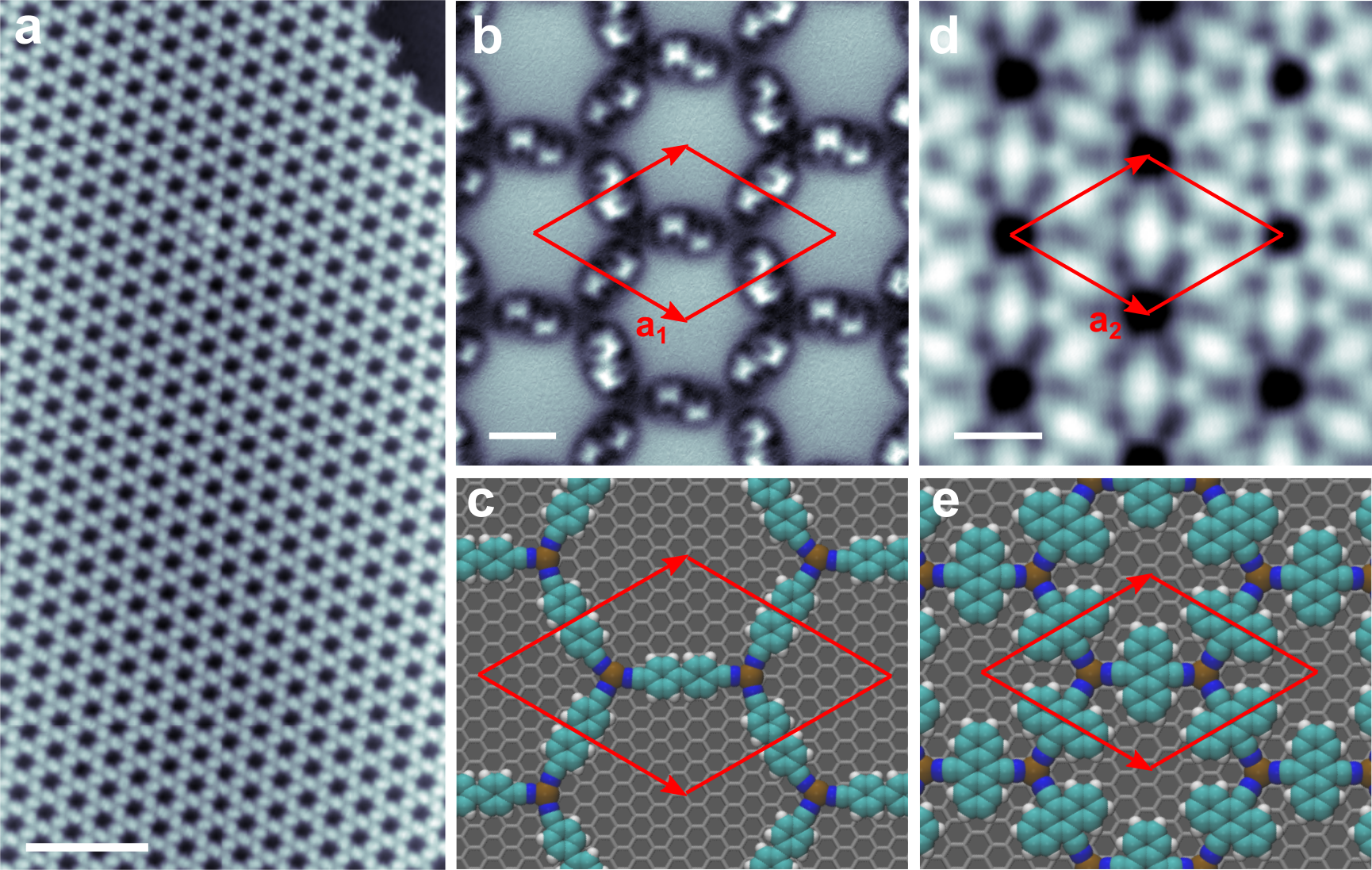}
  \caption{\textbf{Overview of two MOFs.} \textbf{a}, An STM overview image of a honeycomb  DCBP$_3$Co$_2$ MOF on G/Ir(111) surface. Scale bar is 10 nm, imaging parameters: 1.23 V and 3.3 pA. \textbf{b}, Constant height frequency-shift ()$\Delta f$) nc-AFM image of DCBP$_3$Co$_2$ MOF acquired with a CO-terminated tip. Scale bar is 1 nm. \textbf{c}, DFT simulated structure of DCBP$_3$Co$_2$ MOF on graphene. \textbf{d}, STM topography image of DCA$_3$Co$_2$ MOF. The scale bar is 1 nm, imaging parameters: -1 V, 15 pA. \textbf{e}, DFT simulated structure of DCA$_3$Co$_2$ MOF on graphene. Red parallelogram indicates the unit cell.}
  \label{fig:fig1}
\end{figure}

Figure~\ref{fig:fig1} shows the structure of the honeycomb MOFs - DCBP$_3$Co$_2$ and DCA$_3$Co$_2$ - synthesized on epitaxial graphene grown on Ir(111) (G/Ir(111)) surface. %G/Ir(111) surface is the substrate of choice because of its weak geometric ($\sim$50 pm) and work function ($\sim$100 mV) corrugation across the moir\'e unit cell \cite{Hamalainen2013,Altenburg2014}. 
Figure~\ref{fig:fig1}a shows an overview STM topography image of DCBP$_3$Co$_2$ MOF possessing a long-range ordered honeycomb structure. An atomically resolved non-contact AFM (nc-AFM) image of DCBP$_3$Co$_2$ MOF using a CO-terminated tip is shown in Fig.~\ref{fig:fig1}b. The hexagonal symmetry and non-planarity of DCBP molecules (finite torsional angle between two phenyl rings along the long axis) making the framework chiral are readily apparent. This is consistent with the DFT calculated structure on graphene (Fig.~\ref{fig:fig1}c) and simulated nc-AFM image as shown in  Supplementary Information (SI) Fig.~S1. 
%The non-planarity is consistent with earlier reports on similar molecules \cite{Schlickum2008, Palma2015} as well as the DFT optimized structure of gas-phase DCBP molecule and DCBP$_3$Co$_2$ MOF as shown in Fig. ~\ref{fig:fig1}c (see Supplementary Information (SI) for details). 
The calculated gas-phase structure shows that the cobalt atom is in the plane of the framework while it relaxes slightly (about 10 pm) towards the surface on graphene. 

Similar to the DCBP$_3$Co$_2$, DCA$_3$Co$_2$ MOF also reveals a symmetric honeycomb structure as shown by the STM topography image in Fig.~\ref{fig:fig1}d. A typical STM image of a large area DCA$_3$Co$_2$ MOF is shown in SI Fig.~S2 where various domains of different sizes are clearly visible. Compared to DCBP$_3$Co$_2$, the domains of the DCA$_3$Co$_2$ MOF are smaller in size probably due to the more limited mobility of DCA on the graphene surface. A DFT simulated structure corresponding to DCA$_3$Co$_2$ framework is shown in Fig.~\ref{fig:fig1}e. Here also the DCA molecules and cobalt lie in the plane of the framework for gas-phase optimized structures (see SI for the computational details). In both MOFs, we estimate the N-Co coordination bond length from the high resolution STM images to be 1.55 $\pm$0.5 \r{A} (see SI Fig.~S3) which is comparable to the value extracted from DFT relaxed structures and earlier reports \cite{Schlickum2007}. Further, the measured lattice constant of the DCBP$_3$Co$_2$ MOF, \textbf{a$_1$} is 27.9 $\pm$ 0.4 \r{A} while DCA$_3$Co$_2$ MOF possesses a lattice constant \textbf{a$_2$} of 19.6 $\pm$ 0.2 \r{A} (compared to the 27.3 \r{A} and 20.0 \r{A} as extracted from our DFT optimized structure, respectively).  

We synthesize the MOFs by first depositing the organic molecules, followed by deposition of the metal atoms with subsequent annealing. Each of the honeycomb MOFs is separately preceded by the formation of an assembly of single complexes upon deposition of Co atoms on the molecular-layer on G/Ir(111) surface. The network of these complexes is stabilized through intermolecular hydrogen bonds between the cyano and phenyl groups. While DCBP forms four-fold mononuclear single complexes (DCBP$_4$Co) and a stripe of four-fold framework (DCBP$_3$Co) depending on the DCBP:Co stoichiometry, DCA forms only mononuclear three-fold (DCA$_3$Co) complexes which is unambiguously confirmed by nc-AFM image (see SI Figs.~S4 and S5). 
%Similar mononuclear four-fold complexes has been observed for 7,7,8,8-tetracyanoquinodimethane (TCNQ) molecules with Ni and Mn centers \cite{Tseng2011}. 
We attribute the absence of four-fold DCA$_4$Co to a larger steric hindrance compared to that of a four-fold structure of DCBP$_4$Co. 

\begin{figure}[h!]
  \includegraphics[width=0.8\textwidth]{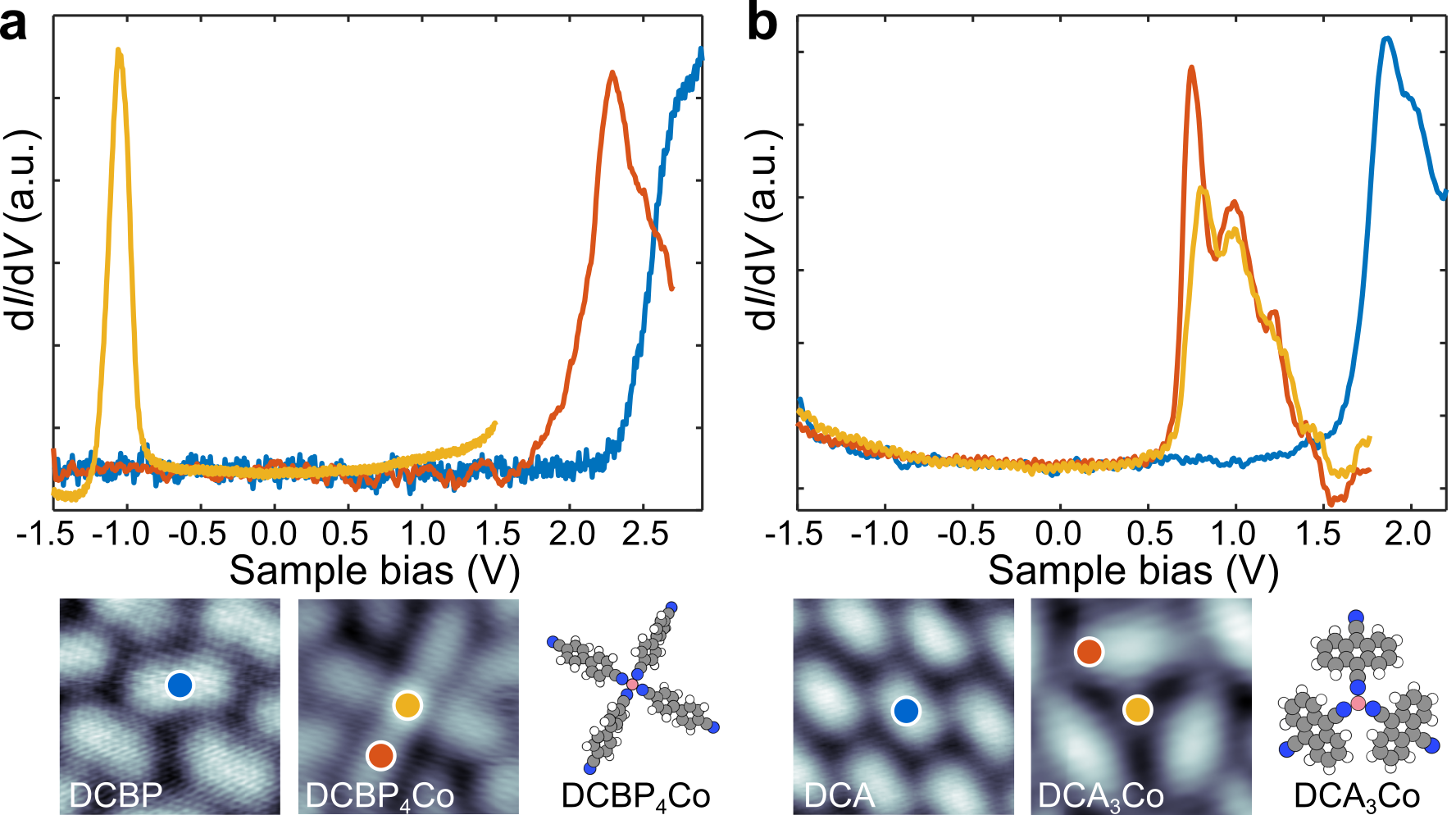}
  \caption{\textbf{STS on single complexes.} \textbf{a}, d$I$/d$V$ spectra  measured on a single DCBP molecule (blue curve), and the Co atom (orange) and on the ligand (red) of a DCBP$_4$Co complex. \textbf{b}, d$I$/d$V$ spectra measured on a single DCA molecule (blue curve), and the cobalt atoms (orange) and the ligand (red) on a single  DCA$_3$Co complex. The positions of the spectra are shown on the bottom panels.}
  \label{fig:fig2}
\end{figure}

%\textbf{Electronic Properties of single complexes--} 
Figure~\ref{fig:fig2} compares d$I$/d$V$ spectra recorded on single molecules and the corresponding single metal-organic complexes. As shown in  Fig.~\ref{fig:fig2}a, d$I$/d$V$ spectrum recorded on a DCBP molecule shows a shoulder at 2.7 V corresponding to the lowest unoccupied molecular orbital (LUMO) (see SI Fig.~S5). The peak due to the highest occupied molecular orbital (HOMO) of the molecule is not visible within the recorded bias range of the spectrum. On a single DCBP$_4$Co complex, d$I$/d$V$ reveals two peaks at 2.3 V and -1 V with corresponding electronic states located on DCBP and Co center, respectively. 
%In the four-fold complexes, Co atom is significantly brighter than in the three-fold complexes manifesting into strong peak-height at -1 V. Cobalt in DCBP single complex is more decoupled \PL{Compared to what?} as evident from the NDR effect in the spectra. 
Based on the bias-dependent STM imaging of the four-fold phases and d$I$/d$V$ spectroscopy of DCBP molecule as a function of distance from Co center (see SI Figs.~S5 and S6), it is clear that the peak 2.3 V originates from the LUMO of the DCBP molecule. The shift of the molecular LUMO towards the Fermi level by 0.4 V indicates that there is an electrostatic shift of the orbital energy due to the Co atom of the complex and other complexes present in the vicinity. 
%Similar gating effects has been observed in Au-coordinated organic complexes by Yang et al. \cite{Yang2014}. This gating is further enhanced in the stripe-phase as the concentration of Co atoms per unit area is higher in the stripe phase than in the single complexes (see SI). There is a comparable shift in the metal-based states in the stripe phase (peak at -1.2 V) compared to single complexes. The lower panels in Fig~\ref{fig:fig2}a shows that the bias-dependent STM imaging of single complex and the stripe-phase resemble strongly the DFT calculated images. 

d$I$/d$V$ spectra on single DCA molecule on G/Ir(111) also reveals a peak at 1.8 V as shown in Fig~\ref{fig:fig2}b. 
%The STM image recorded at 0.4 V and 1.6 V resembles the backbone and LUMO orbital of the neutral molecule as calculated from DFT, respectively (see SI). 
The gating effect due to Co atoms is also observed in the DCA$_3$Co single complexes. The d$I$/d$V$ spectra recorded on DCA of the complex shows that the LUMO shifts down to 750 mV and three satellite vibronic peaks also become visible. The vibronic mode energy of $\sim$200 mV fits well with the expected energy of the C-C vibration \cite{Repp:2010NatPhys,vdLit:2013NatComm}. The assignment of the peak to the molecular LUMO is also evident from the STM images (see SI Figs.~S4 and S7). 
%\PL{Is this sentence needed: Any charge transfer to DCBP or DCA molecules in the respective complexes is ruled out as we do not see any splitting of the LUMO peak across the Fermi energy \cite{Kumar2017}}. 
d$I$/d$V$ spectra recorded on Co center of the complex shows an additional shoulder at the onset of the peak. We attribute this shoulder to the metal-state as the metal center becomes brighter in the STM images at sample bias beyond 0.7 V (see SI Fig.~S4). 
%For the DCA single complexes, the lower panels of Fig.~\ref{fig:fig2}b shows that the bias-dependent STM imaging is in a good agreement with the DFT calculated STM images. 

\begin{figure}
  \includegraphics[width=0.9\textwidth]{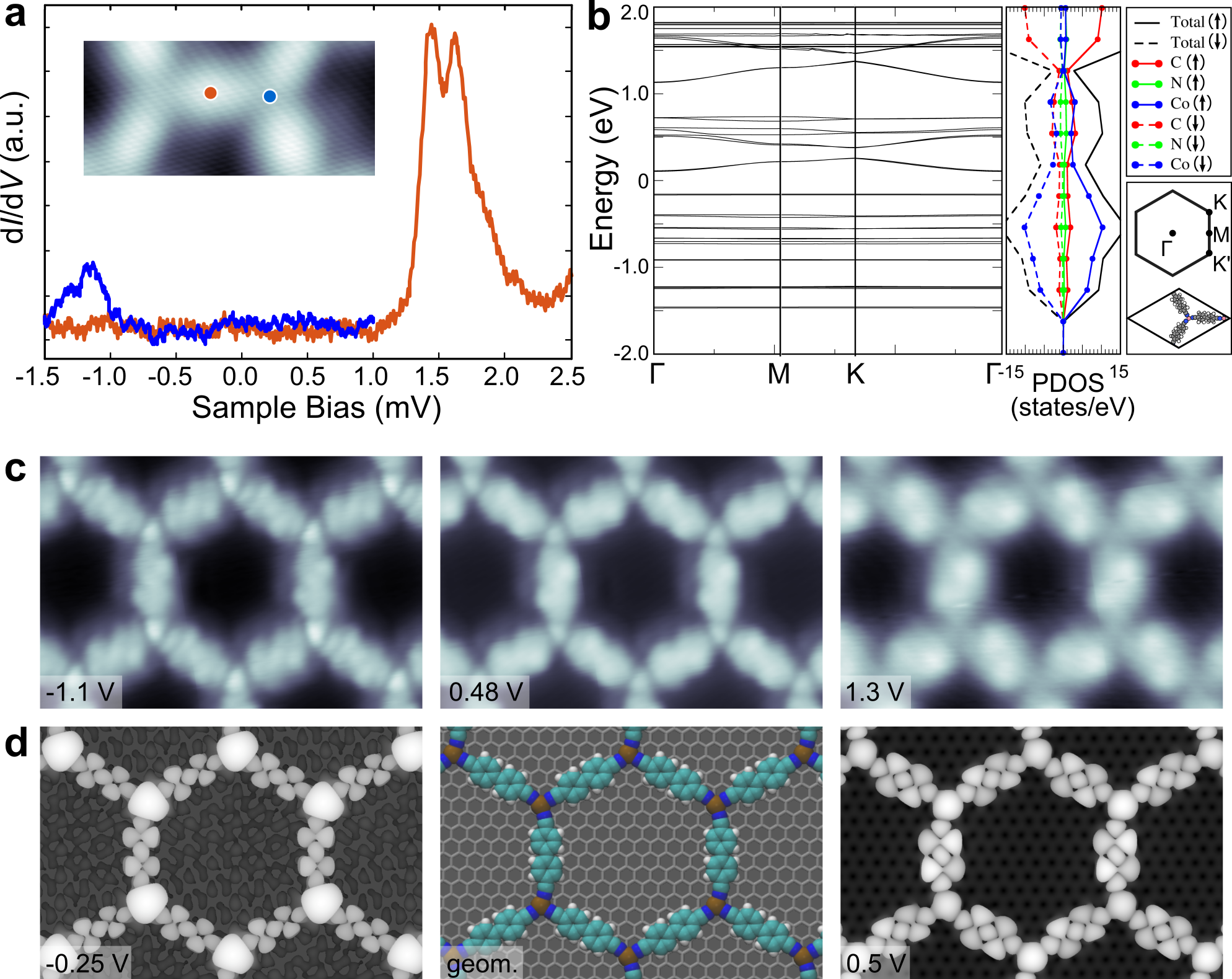}
  \caption{\textbf{Electronic properties of honeycomb DCBP$_3$Co$_2$ MOF.} \textbf{a}, STS recorded on honeycomb DCBP$_3$Co$_2$ at the positions shown in the inset. \textbf{b}, Calculated band structure and total PDOS of DCBP$_3$Co$_2$ MOF. \textbf{c,d} Experimental (panel c) and simulated STM images (panel d) at the energies indicated in the figure. Scan size is $6.2\times 4$ nm$^2$.}
  \label{fig:fig3}
\end{figure}

%\textbf{Electronic Properties of weakly in-plane hybridized DCBP$_3$Co$_2$ MOF--} 
Figure~\ref{fig:fig3}a shows d$I$/d$V$ spectra recorded on DCBP molecule in the DCBP$_3$Co$_2$ MOF has a peak at 1.44 V which we ascribe to the elastic LUMO peak with corresponding vibronic replica at 1.62 V. The small line-width of molecular LUMO and the observation of satellite vibronic peaks indicates %that the molecule in the honeycomb MOF is strongly decoupled from the substrate. 
%Coordination bonds between a DCBP molecule and two Co centers reduces the coupling between the molecule and the G/Ir(111) substrate \cite{Tseng2011}. 
%It also indicates
that the intermolecular electronic coupling in the framework is weak such that we have isolated molecular electronic states. The spectrum recorded on the Co center reveals a faint peak at -1150 mV, which is visible in the background corrected spectrum (see SI Fig.~S8). The state is localized only at the metal-center.
% and has a relatively larger peak-width compared to that of four-fold complexes. The large peak-width can be due to larger coupling of Co with the underlying substrate. 

We have used DFT to calculate the band structure of the DCBP$_3$Co$_2$ MOF as shown in Fig.~\ref{fig:fig3}b for the antiferromagnetic ground state. While DFT underestimates the band gap, it correctly captures the nature of the lowest lying bands: the occupied states have a stronger metal character compared to the unoccupied states, which are mostly composed of the ligand states (Fig.~\ref{fig:fig3}c,d). The enhanced contrast on the metal atoms and ligands can be seen at negative and positive bias, respectively, compared to the STM topography in the gap (Figs. \ref{fig:fig3}c middle panel). However, DFT seems to overestimate the band-width of the unoccupied ligand-derived states compared to the experiment. This could be related to how well the  torsional angle between the phenyl rings of DCBP molecule is estimated by DFT as this is known to control the $\pi-\pi$ conjugation within the backbone of the molecule \cite{Venkataraman2006}. The coupling is  enhanced for the planar, smaller DCA linker as demonstrated below. 

%The right panel displays the total projected DOS around the Fermi level discerning the contributions from the metal and molecular part. In Figs.~\ref{fig:fig3}c-j, we compare DFT simulated STM images to the high-resolution experimental STM images at various biases. The higher spatial resolution of the orbital has been achieved by recording STM images with molecule modified tip \cite{Repp:2005PRL}. As shown in Fig.~\ref{fig:fig3}i, the STM topography image at bias +1300 mV closely resembles to the DFT simulated STM image at \AK{xx V}. Further, STM topography image at lower bias, e.g. -475 and 475 mV shows LUMO of the DCBP molecules (due to the lorentzian shape of the LUMO peak). Also, non-planarity of the molecules is captured in STM images as well as DFT simulated STM images. 

\begin{figure}[t!]
  \includegraphics[width=0.85\textwidth]{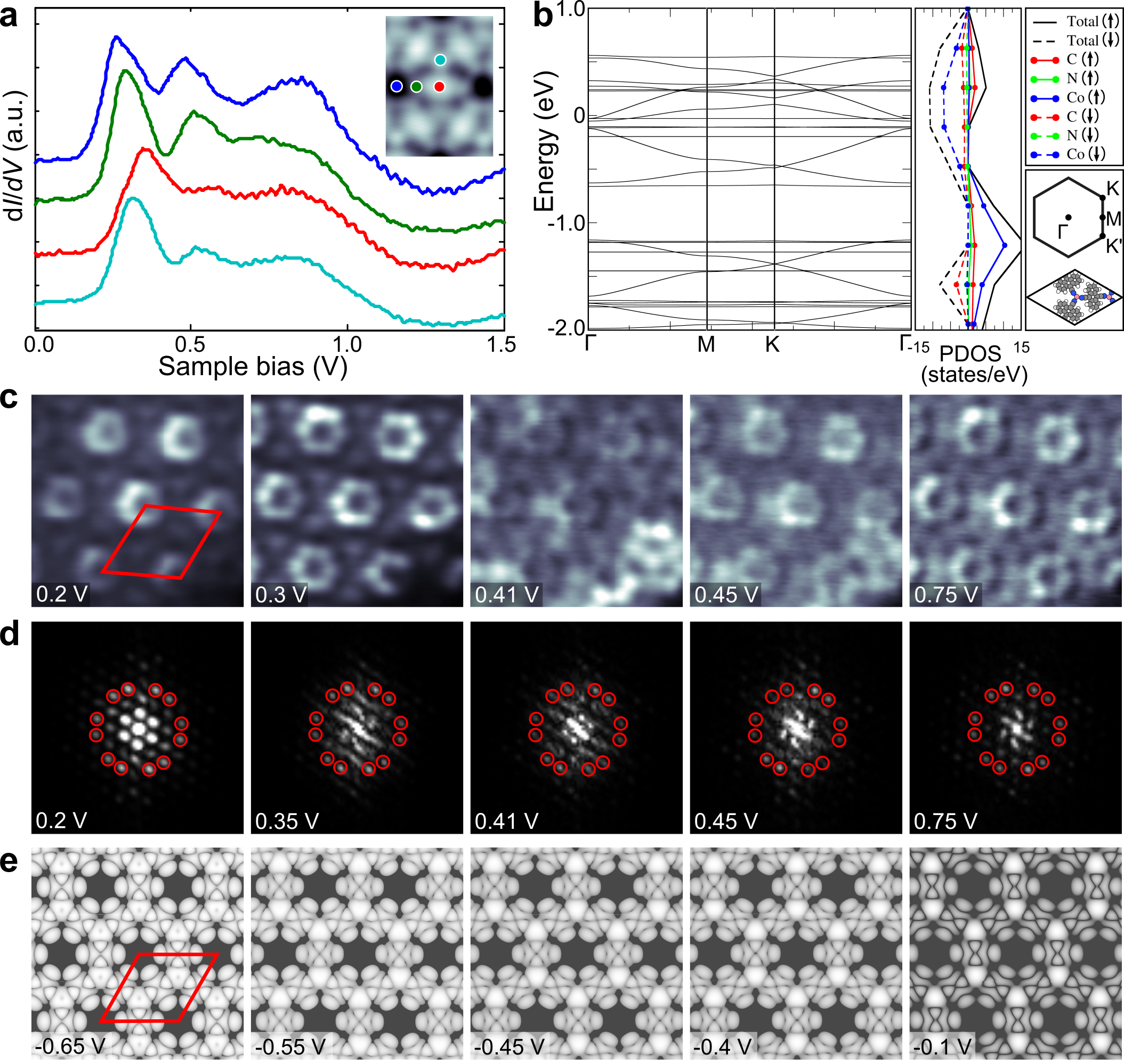}
  \caption{\textbf{Electronic properties of honeycomb DCA$_3$Co$_2$ MOF.} \textbf{a}, STS recorded on honeycomb DCA$_3$Co$_2$ MOF on the positions indicated in the inset. \textbf{b}, Calculated band structure and total PDOS of DCA$_3$Co$_2$ MOF. \textbf{c, d, e}, Experimentally recorded constant-height d$I$/d$V$ (panel c scan size $4.7\times 4.7$ nm$^2$), FFTs of large area d$I$/d$V$ maps (panel d), and simulated LDOS (panel e) at the energies indicated in the panels. The unit cell is indicated by the red parallelogram in panels c and e}
  \label{fig:fig4}
\end{figure}

%\textbf{Electronic Properties of strongly in-plane hybridized DCA$_3$Co$_2$ MOF--} 
%The electronic properties of DCA$_3$Co$_2$ MOF indicates a substantial in-plane electronic hybridization and formation of energy bands with significant  width as evidenced by d$I$/d$V$ spectroscopy, d$I$/d$V$ maps and DFT calculations. 

A substantial in-plane electronic hybridization and formation of energy bands with significant width in DCA$_3$Co$_2$ MOF is evidenced by d$I$/d$V$ spectroscopy, d$I$/d$V$ maps and DFT calculations. 
d$I$/d$V$ spectrum (Fig.~\ref{fig:fig4}, blue) recorded at the center of the ring (constituting six DCA lobes) has three peaks at 260, 480, and 860 mV. Considering the separation between the first and the second peak, $\sim$220 mV, the second peak could still be interpreted as a vibronic satellite. However, the separation between the second and the third peak rules out vibronic origin. The spectrum at the center of DCA molecule (red curve) shows that the first peak shifts to 360 mV, while that on cobalt (cyan curve) has the first peak at 320 mV. The systematic evolution of the spectra across the framework is shown in the SI Fig.~S9. Comparison of the spectra at the lobe and center of DCA and cobalt in the DCA$_3$Co$_2$ MOF to that of DCA$_3$Co single complex indicates that there is additional intensity in the MOF at energies higher than 700 mV (see SI Fig.~S10). As the d$I$/d$V$ signal is directly proportional to the local density of states (LDOS), this is direct evidence of additional electronic states. Further, we have recorded d$I$/d$V$ maps of the same area at different energies as shown in Fig~\ref{fig:fig4}c. At lower energies, 200-300 mV, DCA molecules feature LUMO lobes and Co appears to be bright while at higher energies, e.g. 750 mV, metal states have lower intensity. At intermediate energies, 410 and 450 mV, there exists extra features in the d$I$/d$V$ maps superimposed on the existing framework. To understand these features, we have taken Fourier transforms (FFT) of a large area d$I$/d$V$ maps (Fig. \ref{fig:fig4}d). Apart from the 12 outer spots (red circles) corresponding to the honeycomb structure of DCA$_3$Co$_2$ MOF, there exists internal structure which evolves continuously with the bias. The spots in this quasi-particle interference pattern correspond to scattering vectors connecting the initial and final states of the scattering process at the given energy. In addition to this joint density-of-states, they contain information on the nature of the allowed scattering processes \cite{Petersen:2000FT-STM,Hoffman:2002QPI,Roushan:2009topo}. While quantitative analysis of the experimental patterns is difficult due to the other overlapping peaks stemming from the geometry as well as the limited sample size (number of repetitive unit cells), they indicate the formation of an extended electron system with considerable dispersion (band width). 

The calculated electronic band structure using DFT for the symmetric, ferromagnetic DCA$_3$Co$_2$ framework without graphene is shown in Fig.~\ref{fig:fig4}b. In line with the calculations done for DCA$_3$Cu$_2$ and DCA$_3$Mn$_2$ MOFs \cite{Zhang2016,Wang:2017APL}, the band structure of DCA$_3$Co$_2$ MOF has a number of flat-bands and Dirac cones. While the antiferromagnetic structure is slightly lower in energy (by 0.05 eV), the ferromagnetic state better reproduces the experimental results and we concentrate on it here (see SI Fig.~S11). The presence of a large gap between -0.7 eV and -1.2 eV in the calculated band structure and lack of states below the Fermi energy in the d$I$/d$V$ spectra until -1.5 V suggests that the energy corresponding to the experimental Fermi level lies below the flat band at energy -0.7 V.  %Using the same rationalization for finding Fermi energy for symmetric, antiferromagnetic ground state of DCA$_3$Co$_2$ framework, the electronic band possess no Dirac cones. A comparison of simulated LDOS maps of ferromagnetic and antiferromagnetic structures have been presented in the respective lower panels of measured d$I$/d$V$ maps. A comprehensive list of the simulated maps has been presented in the SI. \PL{Need something here: Apparently, it is difficult to unambiguously determine the magnetic ground state of the DCA$_3$Co$_2$ MOF from simulated LDOS maps}. 
The DFT calculation suggests that the bottom of the conduction band consists of a flat band and a Dirac cone stemming from the DCA states and the Kagome symmetry of the lattice. Subsequently, at higher energies, there are also relatively flat bands originating mostly from the metal atom orbitals and band with more mixed character. This overall picture is consistent with the experiments where we first see intensity on the molecules with metal states emerging at higher energies and an overall band width of $\sim 1$ eV. 

In summary, we have demonstrated synthesis of long range, ordered domains of two  honeycomb MOFs on epitaxial graphene surface. While DCBP$_3$Co$_2$-MOF only has weak coupling between the building blocks, DCA$_3$Co$_2$-MOF shows significant in-plane hybridization resulting in the formation of 2D electronic states with significant band width. These observations point towards the experimental realization of engineered 2D-MOFs with exotic electronic properties. 

\section*{Acknowledgements}

This research made use of the Aalto Nanomicroscopy Center (Aalto NMC) facilities and was supported by the European Research Council (ERC-2011-StG No. 278698 "PRECISE-NANO") and the Academy of Finland (Projects no. 305635 and 311012, and Centres of Excellence Program projects no. 284594 and 284621).

\section*{Author Contributions}
A.K., K.B. and P.L. conceived and planned the experiment. A.K. and K.B. performed the measurements and analysed the data. A.S.F. carried out the DFT calculations. A.K. and P.L. wrote the manuscript and all authors jointly authored, commented, and corrected the manuscript.

\bibliography{achemso-demo}

\newpage

\renewcommand\thefigure{S\arabic{figure}} 
\setcounter{figure}{0}
	
\section*{Supplementary Information}
	
	\noindent
	\large
	\textbf{Simulated nc-AFM image of DCBP$_3$Co$_2$ MOF}
	
	\begin{figure}[h]
		\includegraphics[width=0.7\textwidth]{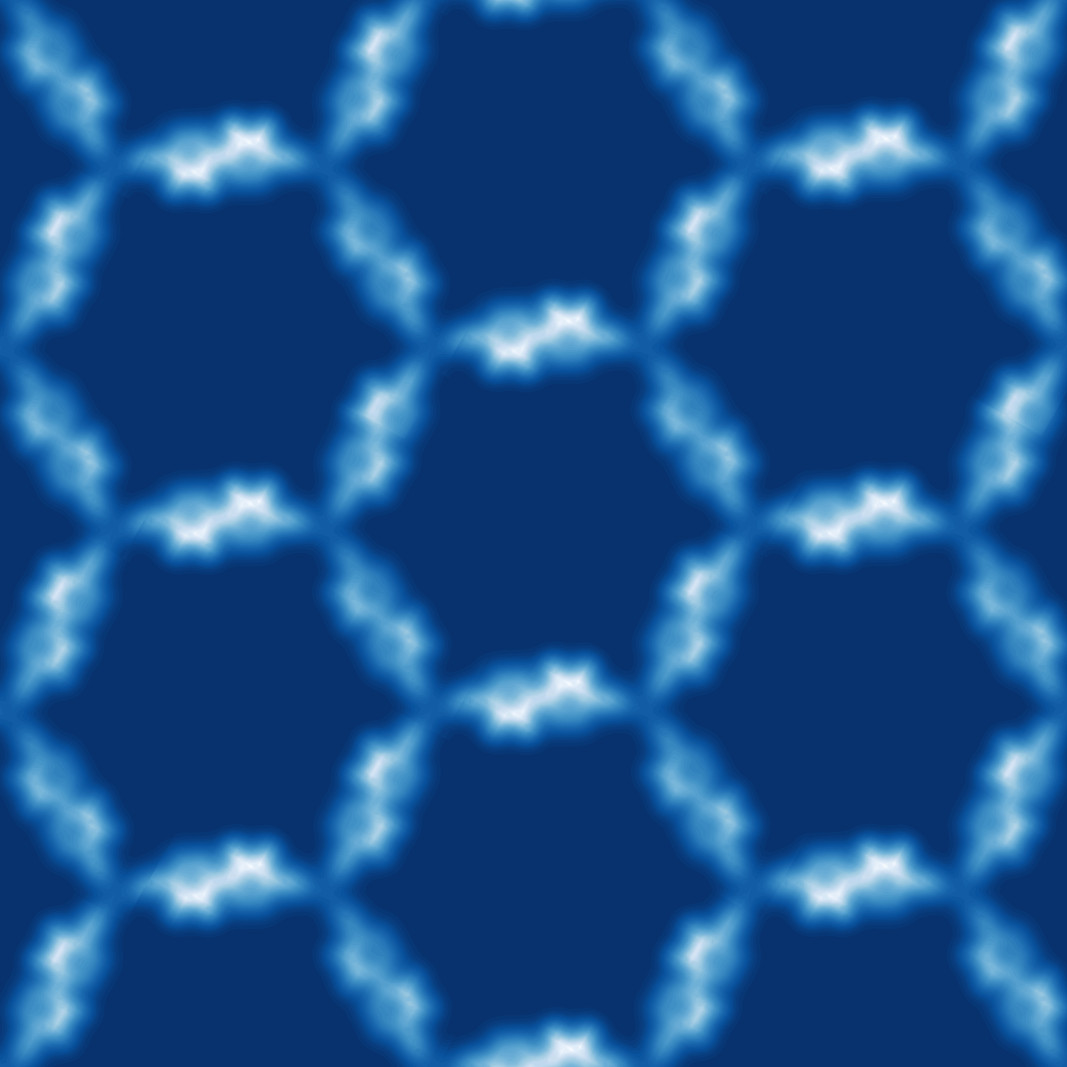}
		\caption{Simulated nc-AFM image of DFT optimized DCBP$_3$Co$_2$ MOF shows non-planarity of DCBP molecules and the chirality exhibited by the framework. }
		\label{fig:af_dca}
	\end{figure}
	
	\renewcommand\thepage{S\arabic{page}} 
	\setcounter{page}{1}
	\newpage

	\textbf{Large area image of DCA$_3$Co$_2$ MOF}
	
	\begin{figure}[h]
		\includegraphics[width=0.8\textwidth]{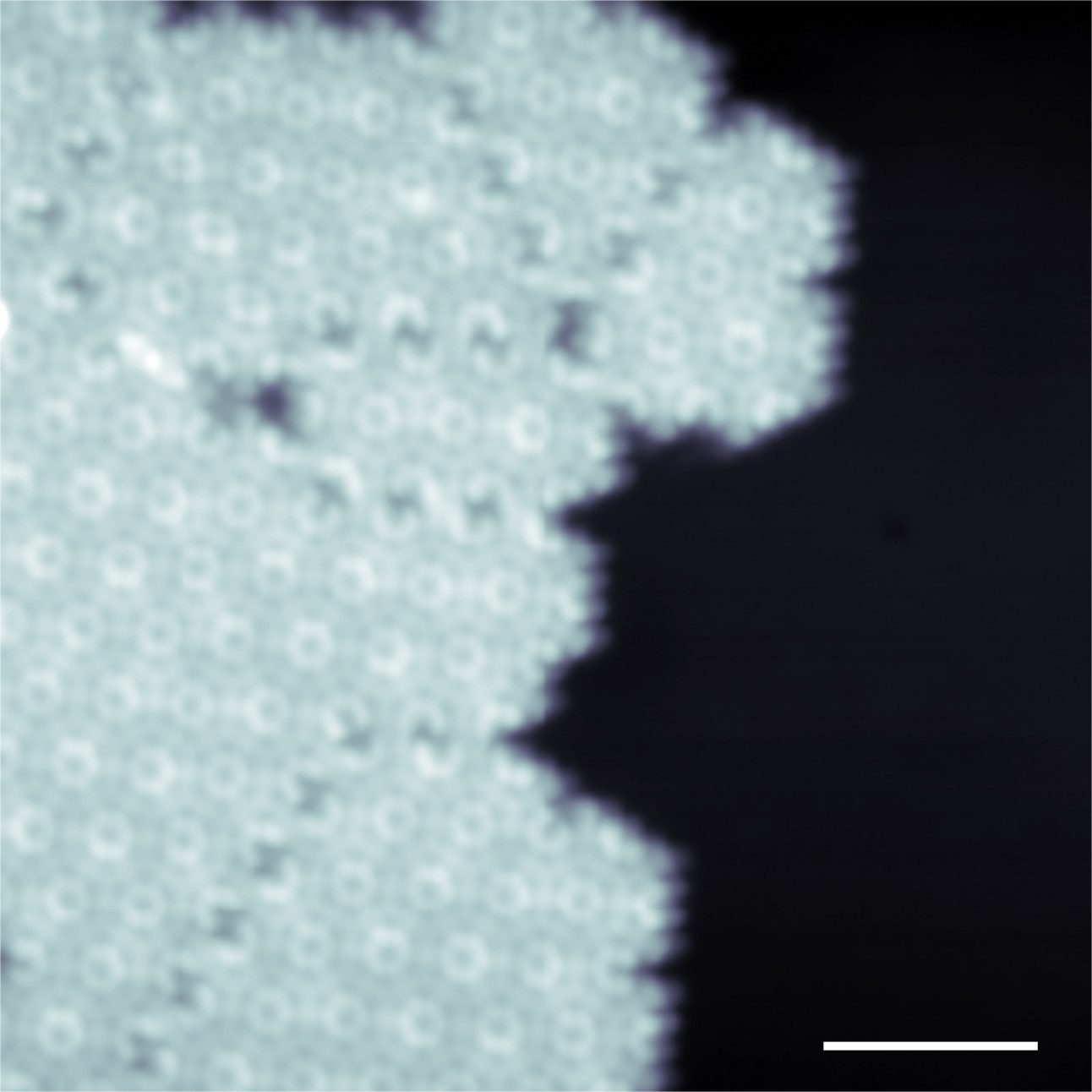}
		\caption{A large area STM image of DCA$_3$Co$_2$ MOF on G/Ir(111) surface (dark area) showing various domains of different sizes. The scale bar is 5 nm and the imaging parameters are $V = 1.3$ V and $I = 20$ pA.}
		\label{fig:si7}
	\end{figure}

	\newpage
	
	\textbf{Co-N bond lengths}
	
	\begin{figure}[h]
		\includegraphics[width=0.9\textwidth]{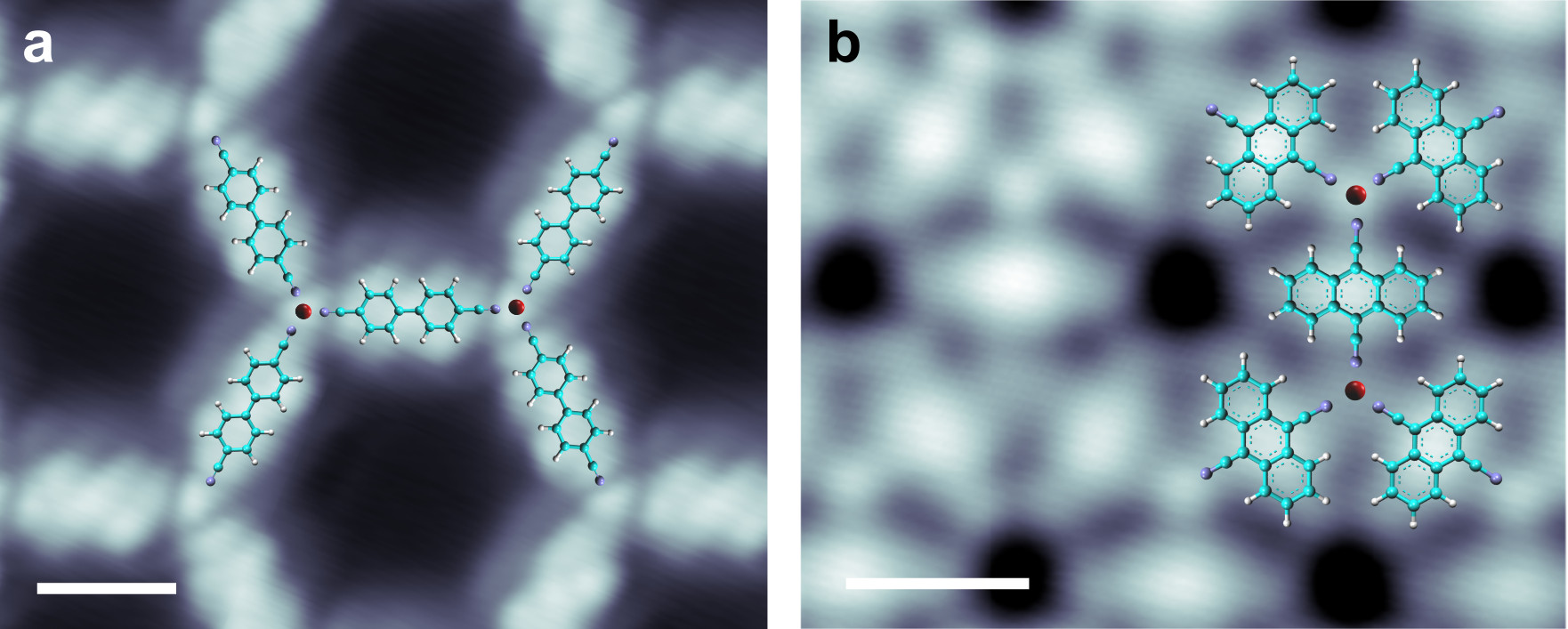}
		\caption{Scaled molecular structures are overlaid on the STM topography image of \textbf{a}, DCBP$_3$Co$_2$ and \textbf{b}, DCA$_3$Co$_2$ MOFs. From here, we extract Co-N bond to be 1.55 $\pm$0.5\r{A}. Scale bars are 10 \AA.}
		\label{fig:CoN_bond}
	\end{figure}

	\newpage
	
	\textbf{Assembly of DCA$_3$Co single complexes on G/Ir(111)}

	\begin{figure}[h]
		\includegraphics[width=0.7\textwidth]{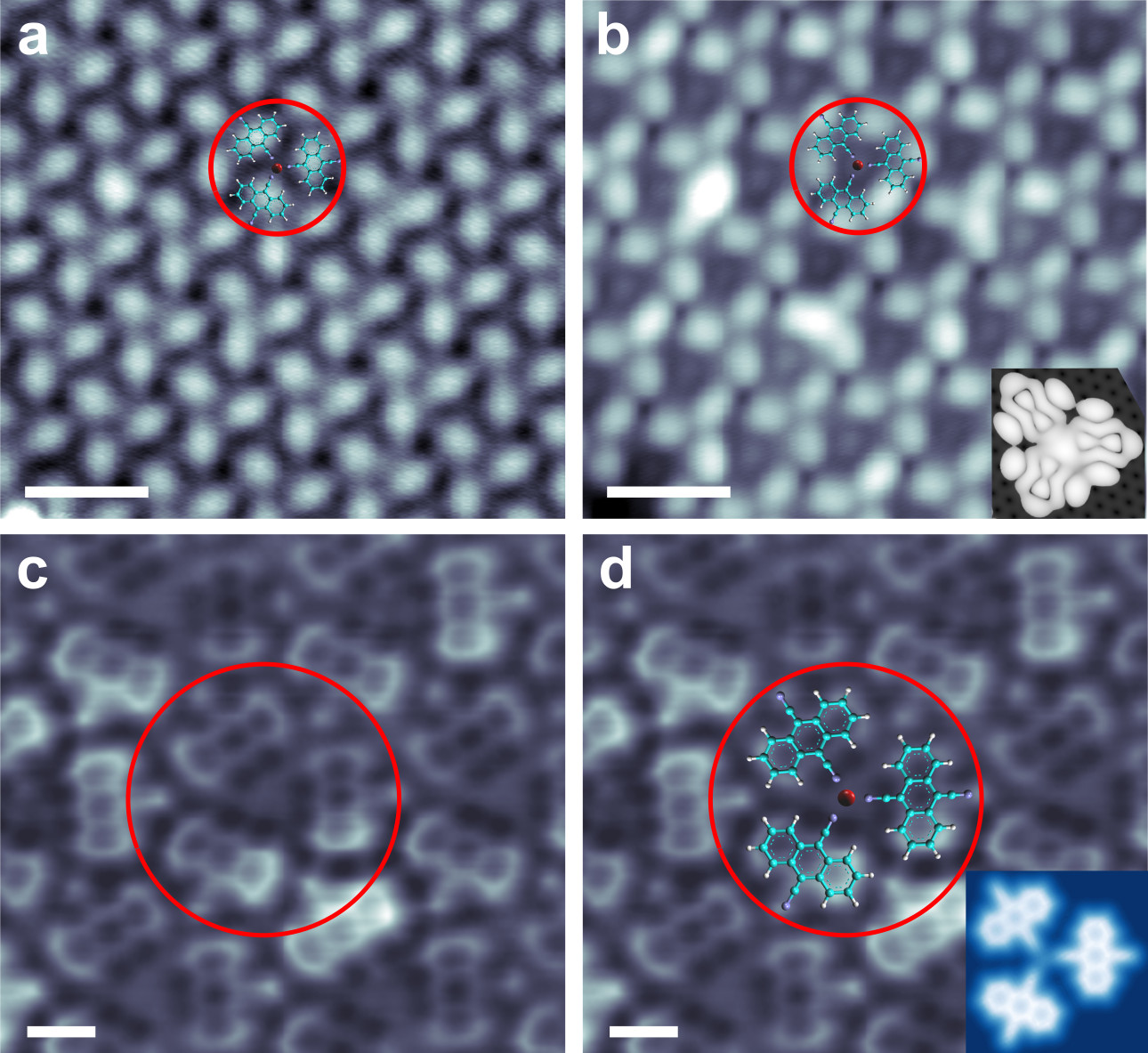}
		\caption{\textbf{a}, A large scale STM image of DCA$_3$Co single complex assembly recorded at bias, $V = 0.3$ V ($I = 2$ pA) showing the backbone of DCA molecules. The red circle indicates a single complex on the surface. The scale bar is 2 nm. \textbf{b}, An STM image of the same area at bias $V = 0.8$ V ($I = 2$ pA) shows LUMO of the DCA molecules and cobalt at the center of the complex. The red circle indicates the same complex indicated in panel a. The inset shows DFT simulated STM image of DCA$_3$Co complex showing LUMO. \textbf{c}, An nc-AFM image of a zoomed-in area shows internal structure of the molecules and their arrangement in the single complex assembly. Again, the red circle indicates a single complex. The cyano groups bonded to cobalt atom are lower than the non-bonded cyano groups. Cobalt atom is not visible at this tip height. \textbf{d}, Same as panel c with an overlaid chemical structure of the DCA$_3$Co single complex. The inset shows a simulated nc-AFM image of the single complex.}
		\label{fig:si5}
	\end{figure}
	
	\newpage
	
	\noindent
	\textbf{Bias dependent STM imaging of DCBP molecule, DCBP$_4$Co single complex, and DCBP$_3$Co stripe-phase}

	\begin{figure}[h]
		\includegraphics[width=0.9\textwidth]{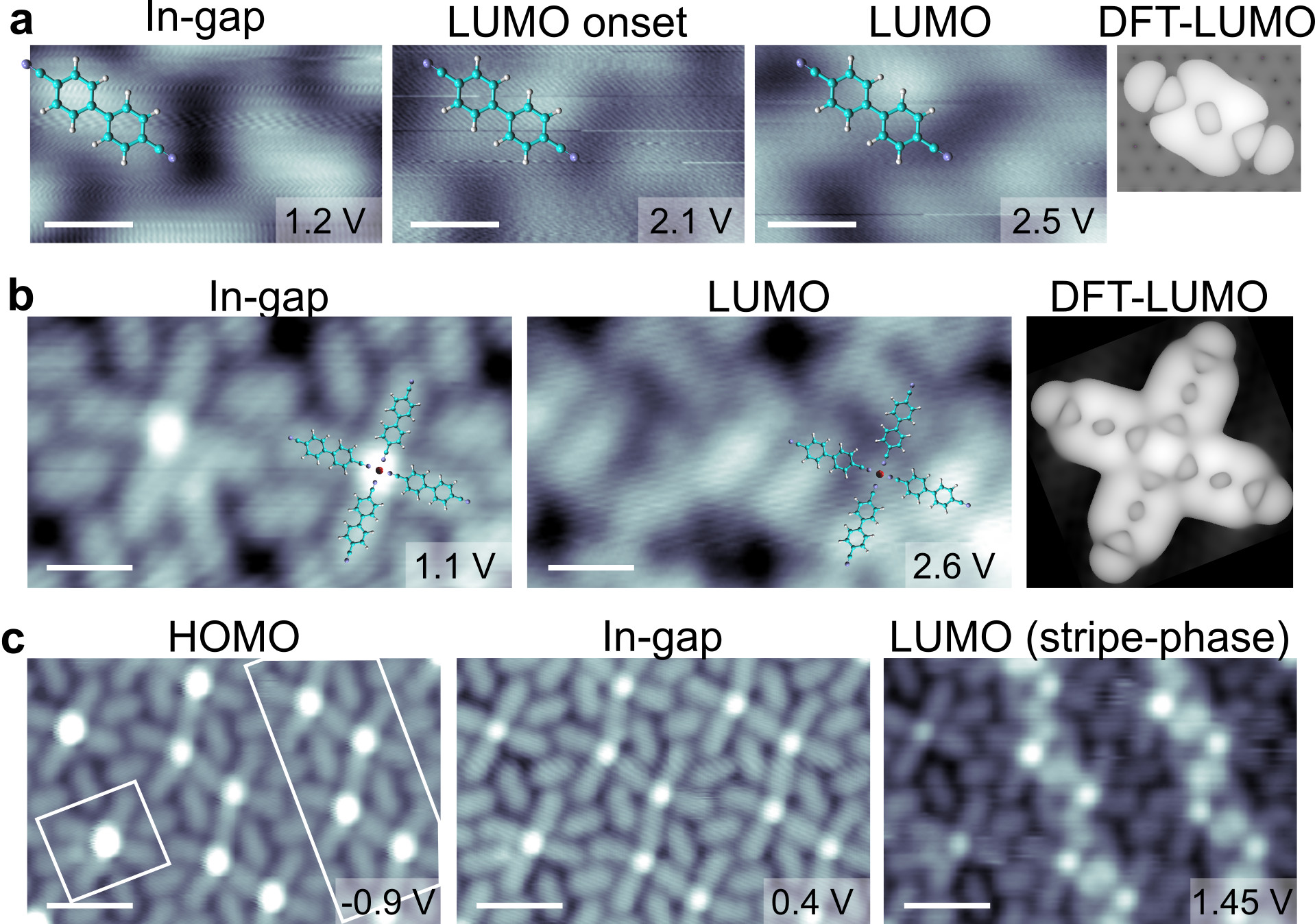}
		\caption{\textbf{a}, Bias dependent STM topography images of DCBP molecules showing backbone (1.2 V), LUMO onset (2.1 V) and LUMO (2.5 V) and DFT simulated STM image showing LUMO. Scale bars are 5 \r{A}. \textbf{b}, Bias dependent STM image of DCBP$_4$Co single complex showing backbone (1.1 V) and LUMO (2.6 V) and DFT simulated STM image depicting LUMO. Scale bars are 1 nm. \textbf{c}, Bias dependent STM images of DCBP$_4$Co single complex and DCBP$_3$Co stripe showing HOMO (-0.9 V) with a bright cobalt center, in-gap image (0.4 V), and LUMO of DCBP$_3$Co stripe-phase (1.45 V). Scale bars are 2 nm.}
		\label{fig:si11}
	\end{figure}
	
	\newpage
	
	\textbf{Distance dependent gating on DCBP LUMO due to cobalt atoms}
	
	\begin{figure}[h]
		\includegraphics[width=0.7\textwidth]{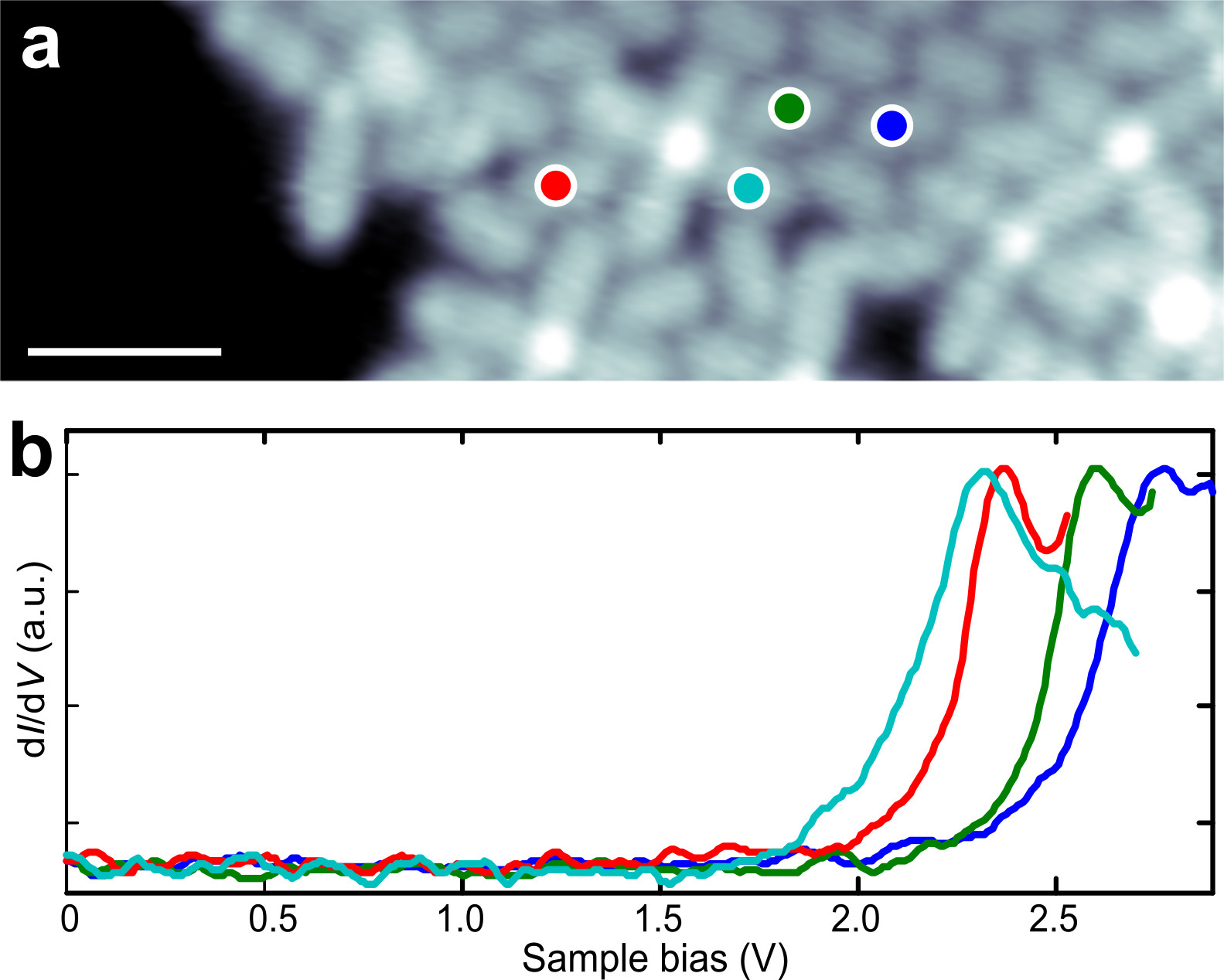}
		\caption{\textbf{a}, STM image of a mixed island of DCPB molecules and DCPB$_4$Co complexes. The scale bar is 2 nm. Sample bias, $V = 0.77$ V and setpoint, $I = 0.7$ pA. \textbf{b}, Gating effect on the DCPB LUMO due to the presence of cobalt metal atoms on the surface. d$I$/d$V$ spectra are compared for DCBP molecules lying at various distances from the cobalt metal atoms: DCBP molecule far away from the cobalt atoms has LUMO peak at 2.75 V (blue curve) which is very close to the LUMO peak in close-packed molecular assembly. The LUMO peak shifts to 2.60 V (green curve) for a DCBP molecule close to a DCBP$_4$Co single complex. For another non-bonded DCBP molecule, close to two single complexes, the peak shifts to 2.37 V (red curve). A representative spectrum (cyan curve) on DCPB molecule bonded to a DCBP$_4$Co single complex has a LUMO peak at 2.30 V.}
		\label{fig:si13}
	\end{figure}
	
	\newpage
	
	\textbf{Assembly of DCA molecules on G/Ir(111)}
	
	\begin{figure}[h]
		\includegraphics[width=0.8\textwidth]{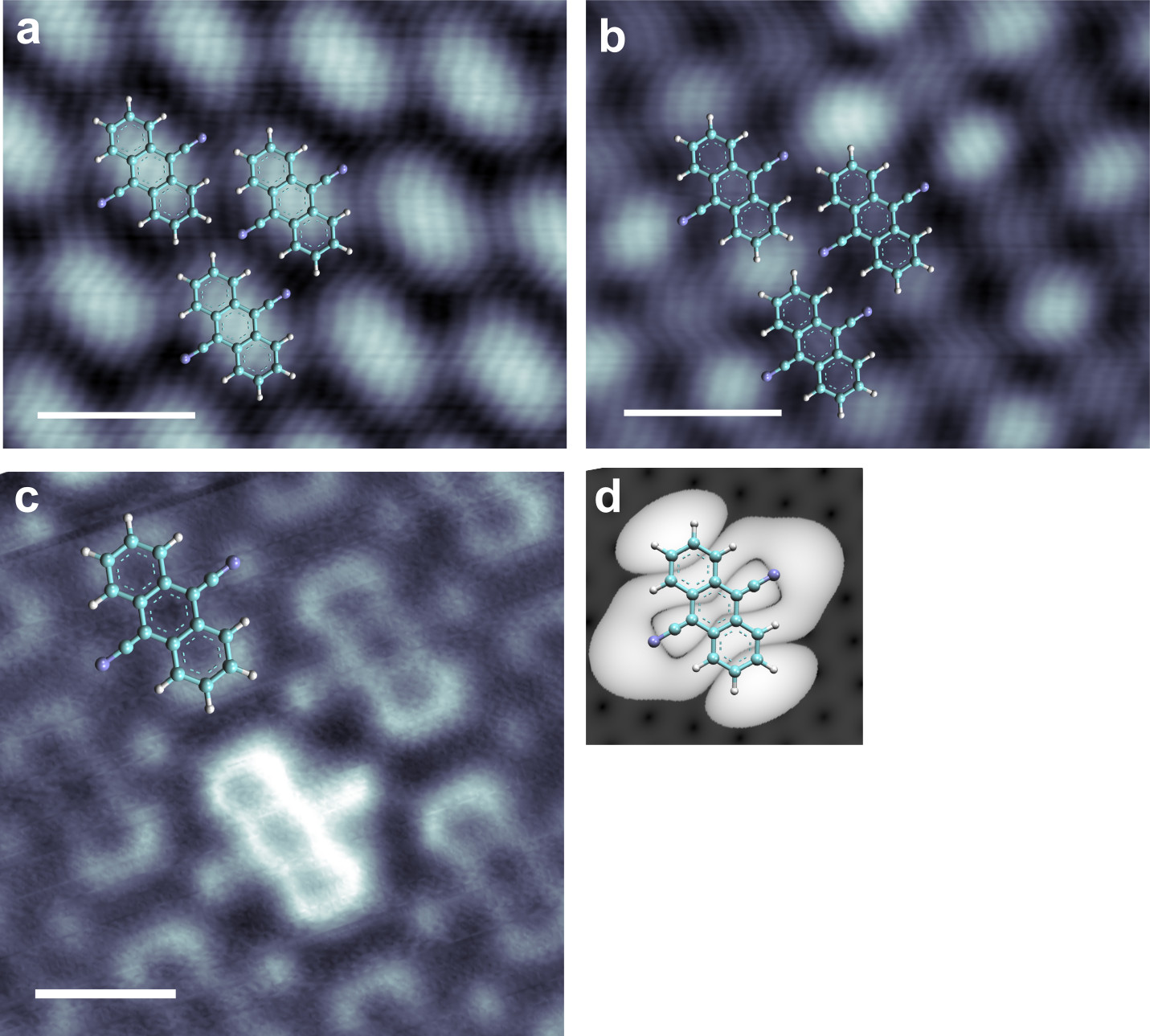}
		\caption{DCA molecule forms close-packed assembly on G/Ir(111) surface. \textbf{a}, STM topography image of DCA molecules at in-gap sample bias V = 0.4 V (I = 1.5 pA) shows molecular backbone. The scale bar is 1 nm. \textbf{b}, STM topography image recorded at 1.6 V (I = 1.5 pA) shows LUMO of the DCA. The scale bar is 1 nm. \textbf{c}, A zoomed-in nc-AFM image of molecular assembly reveals the structure of the DCA molecules and their arrangement. The scale bar is 5 \r{A}. \textbf{d}, DFT simulated STM image of DCA depicting LUMO. Molecular structures are overlaid on each image for the clarity.}
		\label{fig:si4}
	\end{figure}
	
	\newpage
	
	\textbf{d$I$/d$V$ spectrum on Co in DCBP$_3$Co$_2$ MOF}

	\begin{figure}[h]
		\includegraphics[width=0.6\textwidth]{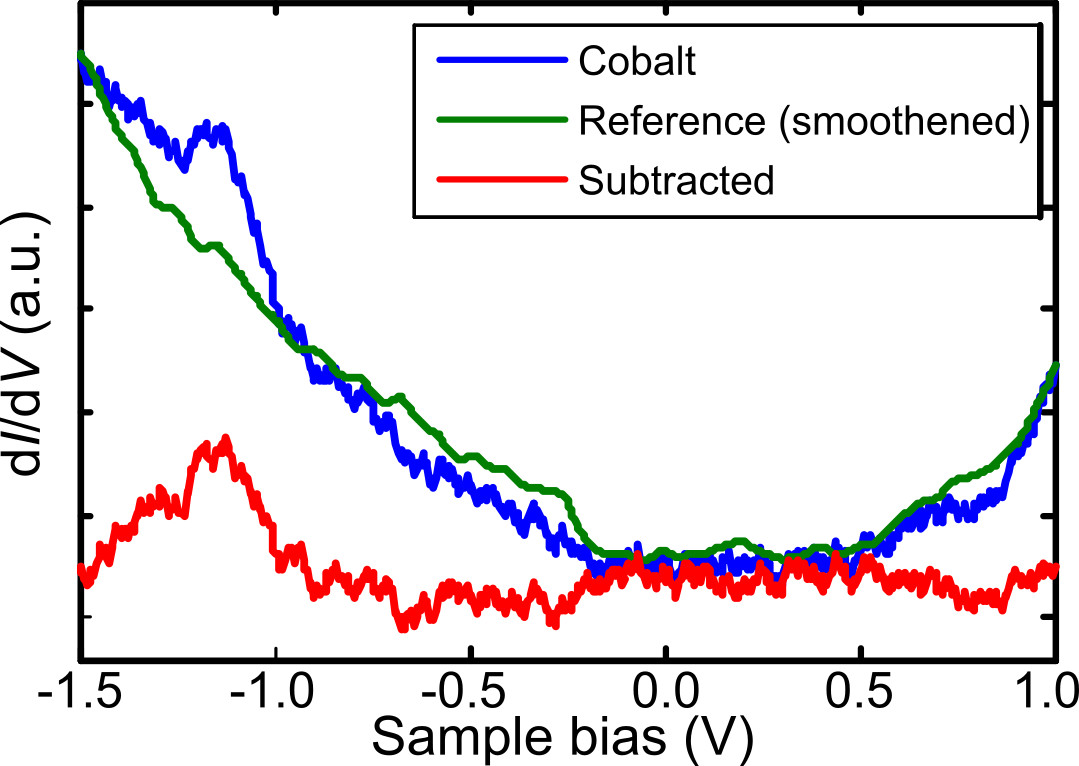}
		\caption{d$I$/d$V$ spectrum (blue curve) on Co in DCBP$_3$Co$_2$ MOF shows a faint peak at -1.15 V. After subtracting the reference d$I$/d$V$ spectrum (green curve) recorded on G/Ir(111) surface, we extract the spectrum on Cobalt (red curve).}
		\label{fig:Co_spec}
	\end{figure}
	
	\newpage
	
	\noindent
	\textbf{High spatial-resolution d$I$/d$V$ spectra on DCA$_3$Co$_2$ honeycomb MOF}
	
	\begin{figure}[h]
		\includegraphics[width=0.8\textwidth]{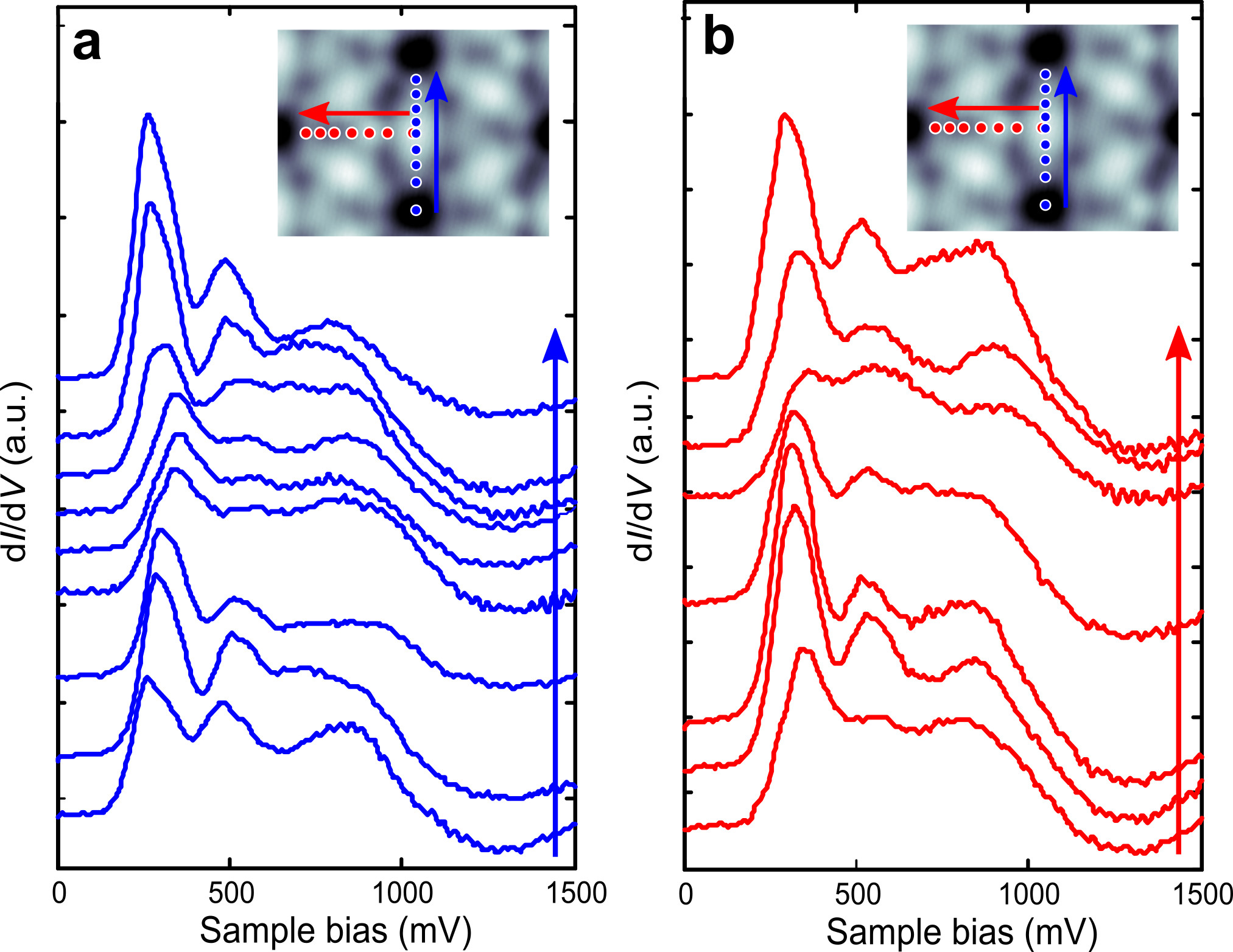}
		\caption{\textbf{a,b} d$I$/d$V$ point spectra recorded along the backbone of DCA (panel a) and across cobalt atom (panel b) of DCA$_3$Co$_2$ MOF as shown in the insets. At the center of the ring of DCA lobes, d$I$/d$V$ spectra shows peak position at 260 mV which shifts to 290 mV at the position of the lobe which further shifts to 350 mV at the center of the molecule. Across the cobalt atom, the peak position shifts to 320 mV from 350 mV at the center of DCA. Inset imaging parameters $V = 120$ mV and setpoint $I = 33$ pA.}
		\label{fig:si1}
	\end{figure}
	
	\newpage
	
	\noindent
	\textbf{d$I$/d$V$ spectra on DCA$_3$Co single complex with DCA$_3$Co$_2$ honeycomb MOF}
	
	\begin{figure}[h]
		\includegraphics[width=0.6\textwidth]{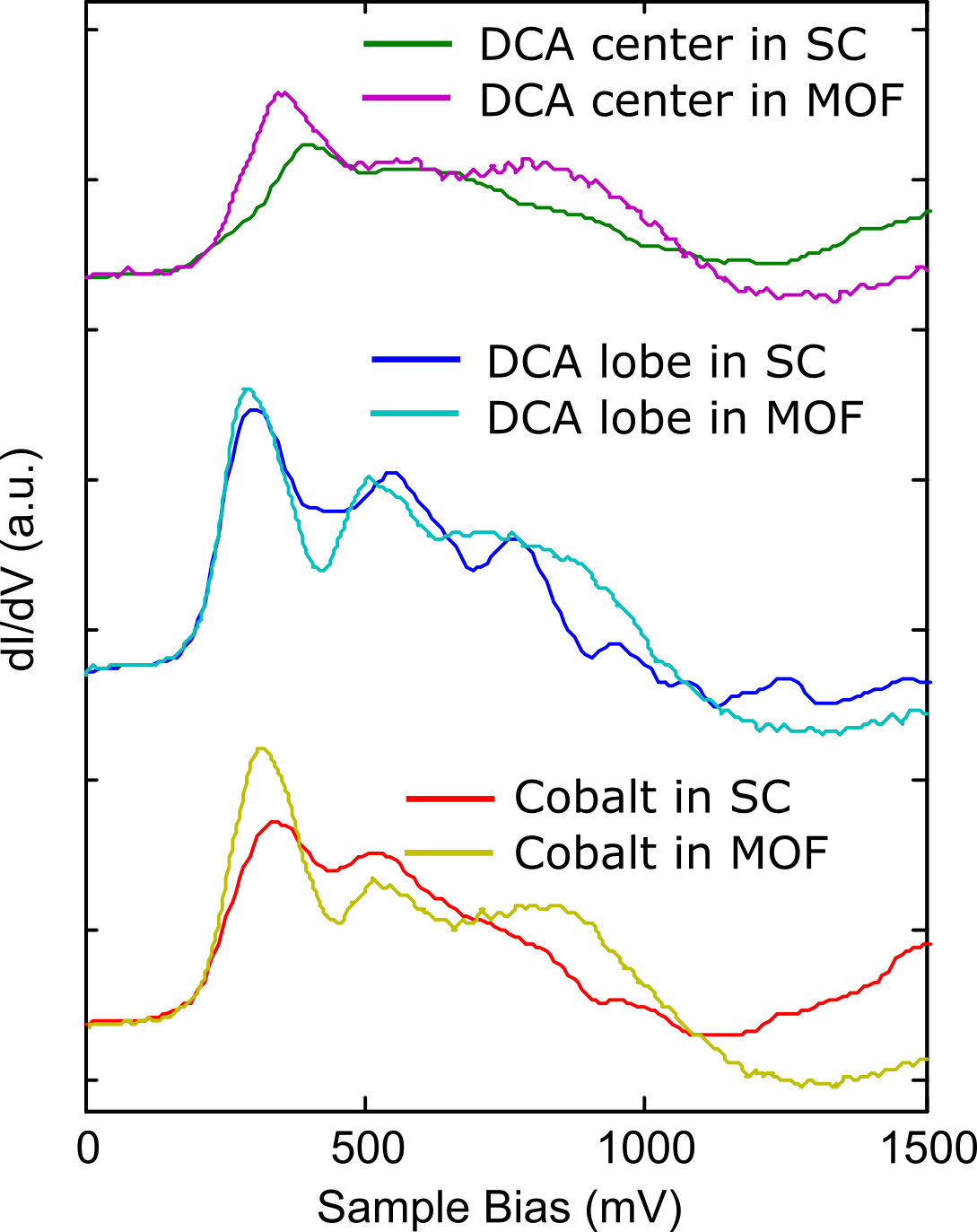}
		\caption{d$I$/d$V$ point spectra comparison between DCA$_3$Co single complex (SC) and DCA$_3$Co$_2$ MOF. The spectra at various locations of MOF are displaced vertically w.r.t each other for visualization and the respective SC spectra are vertically displaced too to match MOF spectra. Further, each SC spectrum is shifted towards fermi energy by 470 mV and multiplied by various factors to achieve approximate normalization. Spectra at DCA center, DCA lobe, and cobalt of SC was multiplied by 2.5, 2.7 and 3.1, respectively. Each spectrum recorded on MOF show enhanced NDR effect, due to further decoupling from the substrate as in DCBP$_3$Co$_2$ MOF. Despite enhanced decoupling, there is at least an excess of states at energies higher than 700 mV. }
		\label{fig:si3}
	\end{figure}
	
	\newpage
	
	\noindent
	\textbf{Band structure and LDOS maps of DCA$_3$Co$_2$ MOF for the antiferromagnetic ground state}
	
	\begin{figure}[h]
		\includegraphics[width=0.8\textwidth]{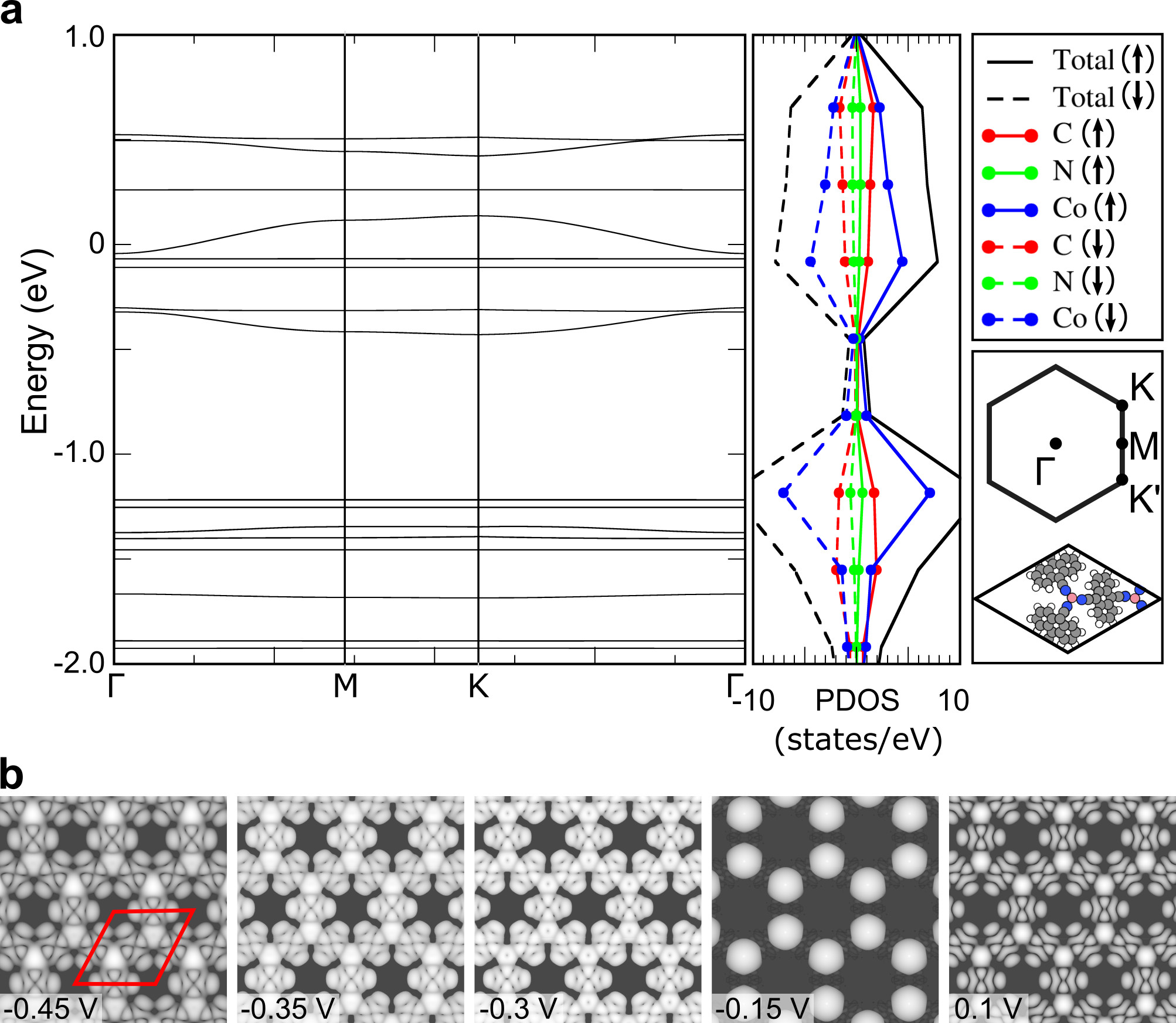}
		\caption{\textbf{a}, Calculated band structure of DCA$_3$Co$_2$ MOF for the antiferromagnetic ground state. The band structure shows a number of gaps at $\Gamma$-point and smaller gaps at K-point which is inconsistent with the measured d$I$/d$V$ spectra. While, the top right panels show the partial density of states, the bottom panel shows the unit cell and corresponding brillouin zone. Presence of a large gap between -0.4 eV and -1.25 eV in the calculated band structure and no observation of states below Fermi energy in the d$I$/d$V$ spectra until -1.5 V suggests that the energy corresponding to the experimental Fermi level will lie below the flat band at energy -0.3 V. \textbf{b}, Simulated LDOS maps for the similar energy range as in measured d$I$/d$V$ band width shows the LDOS map deviates significantly at -0.15 V from the measured d$I$/d$V$ maps. The red rhombus represents the unit cell of the structure.}
		\label{fig:af_dca}
	\end{figure}
	
	\newpage

	\textbf{Methods}
	\normalsize
	
	\emph{Sample preparation.} The experiment was performed in a Createc low-temperature STM/AFM equipped with a preparation chamber with a base pressure lower than $1\times 10^{-10}$ mbar. Ir(111) sample was cleaned by repetitive cycles of sputtering using high energy (2 kV) Ne beam and annealing in oxygen environment at 900 $^\circ$C followed by flashing to 1300 $^\circ$C. Graphene was grown by adsorbing ethylene and flashing the sample to 1100 - 1300 $^\circ$C in a TPG (temperature programmed growth) step followed by a CVD (chemical vapour deposition) step where the Ir(111) substrate at 1100 - 1300 $^\circ$C is exposed to ethylene gas at $4\times 10^{-7}$ mbar pressure for around 60 s. This gives approximately a full monolayer coverage of graphene (G/Ir(111)). 
	
	DCBP$_x$Co$_y$ structures were prepared by sequential deposition of the molecule and cobalt atoms on G/Ir(111) substrate at various temperatures. A submonolayer close-packed assembly of DCBP molecules is achieved by depositing it on G/Ir(111) substrate kept at room temperature using a home build evaporator at 47 $^\circ$C. Addition of cobalt using a high-temperature effusion cell to the molecular assembly leads to a spontaneous formation of DCBP$_4$Co single complexes and DCBP$_3$Co stripe-phase domains depending on the DCBP:Co stoichiometry. To synthesize DCBP$_3$Co$_2$ honeycomb MOF, cobalt was further added to the substrate at temperature 60 - 70 $^\circ$C. After each stage of structure formation, the sample was transferred to the STM and the measurements were performed at 4.5 K. 
	
	Similarly, DCA$_x$Co$_y$ structures were prepared by sequential deposition of the molecule and cobalt atoms on G/Ir(111) substrate at various temperatures. Addition of cobalt to the submonolayer assembly of DCA molecules (evaporation temperature 60 $^\circ$C) on G/Ir(111) substrate at room-temperature leads to spontaneous formation of assembly of DCA$_3$Co single complexes. To synthesize DCA$_3$Co$_2$ honeycomb MOF, cobalt was fruther added to the assembly of DCA molecules on G/Ir(111) substrate at $\sim$85 $^\circ$C. Alternatively, substrate with assembly of DCA$_3$Co single complexes can be annealed at temperature 80-90 $^\circ$C for 45 minutes to form small domains of DCA$_3$Co$_2$ honeycomb MOF. Prolonged annealing at similar temperature increases the domain size. Also, after each stage of structure formation, the sample was transferred to STM and the measurements were performed at 4.5 K. 
	
	\emph{STM and AFM experiments.} Voltage modulations with amplitude of 10 - 15 mV were used for d$I$/d$V$ spectra and maps. Mechanically cut Pt/Ir tips were used for all the STM measurements. Non-contact AFM (nc-AFM) measurements were carried out using a qPlus sensor with resonance frequency $f_0\sim30.7$ kHz, a quality factor $Q\sim10^5$, a spring constant  $k\sim1.8$ kN/m, and an oscillation amplitude of 50 pm. Here, the tips were functionalized by picking up individual CO molecules on a Cu(111) surface as described elsewhere \cite{co_pickup,Gross2009}. nc-AFM images were acquired by measuring the frequency shift of the qPlus sensor while scanning over the area in constant height mode. The sample bias during nc-AFM imaging was kept at less than 5 mV. WSxM \cite{wsxm} and Gwyddion  \cite{gwyddion1,gwyddion2} software were used to process all STM and nc-AFM images. 
	
	\emph{Computational}
	All first principles calculations in this work were performed using the periodic plane-wave basis VASP code~\cite{Kresse1996,Kresse1996a} implementing the spin-polarized density functional theory (DFT). To accurately include van der Waals interactions in this system we used the optB86B+vdW-DF functional,~\cite{Klimes:2010ei,Klimes:2011gw,Bjorkman:2012kj} selected based on previous work showing that it provides a sufficiently accurate description for all subsystems involved in the measurement. Projected augmented wave (PAW) potentials were used to describe the core electrons,~\cite{Blo94} with a kinetic energy cutoff of 550~eV (with \texttt{PREC=accurate}). Systematic $k$-point convergence was checked for all systems, with sampling chosen according to system size. This approach converged the total energy of all the systems to the order of meV. Significantly increased $k$-point sampling with frozen geometries was used as the basis for the band structure calculations. The properties of the bulk graphite, graphene and the isolated molecular structures were carefully checked within this methodology, and excellent agreement was achieved with experiments where possible. For calculations of the network structures on graphene, system sizes were chosen to minimize lattice mismatch and any remaining strain (less than 1\%) was accommodated in the graphene. Note that for calculations of the DCA molecular network on graphene, DFT consistently predicted that the three-fold symmetry seen experimentally would be broken by displacement of the Co closer to two of the nitrogens by about 0.02 \AA. Although this is at the limits of our accuracy, we cannot exclude that in reality something else plays a role in maintaining the symmetry, for example the metal substrate. Since asymmetric structures were never observed experimentally, we constrained the system to be three-fold symmetric. On graphene, the calculations suggest displacement of Co towards the surface (larger than for DCBP$_3$Co$_2$), which is not really seen in the experiment. Symmetric and asymmetric systems had minor differences in their band structures, especially compared to those between antiferromagnetic and ferromagnetic cases. As a further check, we also performed calculations on a DCA$_3$Mn$_2$ framework using the same approach, and found very similar results to those published previously \cite{Zhang2016}.\\
	\indent
	STM images were calculated using the HIVE package \cite{Vanpoucke:2008dk}. For calculated AFM images we used our implementation of the model developed by Hapala \emph{et al.}~\cite{Hapala:2014bo}. The molecular structure was taken from DFT simulations and the electrostatic potential was extracted from the Hartree potential~\cite{Hapala:2014bo,sin_github}. The mechanical AFM model relies on empirical Lennard-Jones parameters, which were taken from the CHARMM force field~\cite{Brooks:2009ew}. The best agreement with experiment was found with a tip lateral spring constant of about 0.5~N/m, similar to values reported in previous studies~\cite{Hapala:2014kr}. All other parameters are the same as intended by Hapala \emph{et al.}, and the simulated AFM scan is performed at a resolution of 5~pm (in all directions), with a force tolerance criterion of $4 \times 10^{-6}$~eV\AA $^{-1}$. The 3D force field is subsequently converted into a frequency shift image~\cite{Welker:2012dw}, using the experimental parameters.\\

	\newpage

\end{document}